\documentclass[10pt]{iopart}
\usepackage{graphicx}
\eqnobysec

\newcommand{\beq}{\begin{equation}}
\newcommand{\beqa}{\begin{eqnarray}}
\newcommand{\eeq}{\end{equation}}
\newcommand{\eeqa}{\end{eqnarray}}

\newcommand{\cum}[1]{\mean{\!\mean{#1}\!}}
\newcommand{\dd}{{\rm d}}
\newcommand{\eps}{\varepsilon}
\newcommand{\erf}{\mathop{\rm erf}}
\newcommand{\erfc}{\mathop{\rm erfc}}
\newcommand{\etc}{\!\cdots\!}

\newcommand{\fras}[2]{{\scriptstyle{\scriptstyle#1\over\scriptstyle#2}}}
\newcommand{\h}[1]{{\widehat{#1}}}
\newcommand{\half}{{\fras{1}{2}}}
\newcommand{\ii}{{\rm i}}
\newcommand{\lam}{\lambda}
\newcommand{\li}{{\{L_i\}}}
\newcommand{\lll}{{L\dots L}}
\newcommand{\mean}[1]{\langle#1\rangle}
\newcommand{\pei}{{\{p_i\}}}
\newcommand{\prob}[1]{{{\rm Prob}\{#1\}}}
\newcommand{\stir}[2]{\left[#1\atop#2\right]}
\newcommand{\G}{{\cal G}}
\newcommand{\Int}[1]{{\lfloor#1\rfloor}}
\newcommand{\Lam}{\Lambda}

\begin{document}

\title{On multidimensional record patterns}

\author{P L Krapivsky$^{1,2}$ and J M Luck$^2$}

\address{$^1$ Department of Physics, Boston University, Boston, MA 02215, USA}

\address{$^2$ Institut de Physique Th\'eorique, Universit\'e Paris-Saclay,
CEA and CNRS, 91191 Gif-sur-Yvette, France}

\begin{abstract}
Multidimensional record patterns are random sets of lattice points
defined by means of a recursive stochastic construction.
The patterns thus generated owe their richness
to the fact that the construction is not based on a total order,
except in one dimension,
where usual records in sequences of independent random variables are recovered.
We derive many exact results on the statistics of multidimensional record patterns
on finite samples drawn on hypercubic lattices in any dimension $D$.
The most detailed analysis concerns the two-dimensional situation,
where we also investigate the distribution of the landing position
of the record point which is closest to the origin.
Asymptotic expressions for
the full distribution and the moments of the number of records on large hypercubic samples
are also obtained.
The latter distribution is related to that of the largest of $D$ standard Gaussian variables.
\end{abstract}

\eads{\mailto{pkrapivsky@gmail.com},\mailto{jean-marc.luck@ipht.fr}}

\maketitle

\section{Introduction}

In this work we introduce and investigate multidimensional record patterns.
These are random sets of lattice points defined by means of a recursive construction,
generalizing the construction of usual records, i.e.,
records in sequences of independent and identically distributed random variables.
The statistics of records has a long history~\cite{chan,ren1,ren2}
and it has been the subject of many books and reviews~\cite{gli,arn,nev,s+z,bun,wer}.
The recent review~\cite{god},
as well as References~\cite{KR02,krug,M1,glrec,mms,LDW,WBK,M3,M8,lmwin,klth},
provide a variety of examples of applications of record statistics to physical problems.

The multidimensional record patterns considered here
are non-trivial extensions of records to an arbitrary higher dimension $D$.
Their recursive construction is entirely parameter-free
and respects all symmetries of the underlying hypercubic lattice.
Their statistics therefore involve some universal combinatorics,
generalizing that involved in the case of usual records.
The present work was originally inspired by recent studies
on the fragmentation of rectangles drawn on the square lattice~\cite{spain,stick},
where an initial rectangle of size $L\times M$ may break either vertically or horizontally
into two smaller rectangles.
In a symmetric variant of the model it breaks simultaneously in both directions,
giving birth to four rectangles.
The process is repeated until it stops when the entire system is filled with frozen rectangles,
called sticks, of size $K\times1$ or $1\times K$.
Several quantities of interest have been studied in~\cite{spain,stick},
including the mean number of sticks and the distribution of the stick length~$K$.
The outcomes provide interesting examples of the non-trivial scaling laws
which are observed in fragmentation processes
(see e.g.~\cite{R18,F1,F2,F3,F4,F5,F6,R20,R22}, and~\cite{KRB} for a review).
The connection with the present work is as follows.
If one monitors the successive breaking events of the symmetric fragmentation process
which are closer and closer to the origin
(i.e., to the lower left corner of the rectangular sample),
one obtains a random set of points which identifies with a two-dimensional record pattern.

There is no direct connection
between the multidimensional record patterns studied here
and records in sequences of random variables,
except in the one-dimensional situation, reviewed in section~\ref{1d}.
First of all, there is no canonical notion of records in more than one dimension.
Consider for definiteness a two-dimensional signal $x_{l,m}$ for $l,m=1,2,\dots$
Along the lines of the usual theory of records,
we say that there is a record at the point $(l,m)$
if~$x_{l,m}$ is larger than $x_{l',m'}$ for all $(l',m')<(l,m)$.
The definition of records therefore relies
on the choice of a total order among lattice points,
i.e., couples of integers~$(l,m)$.
Natural requirements are as follows:
the total order should respect the translation invariance
of the underlying lattice,
and be such that every point $(l,m)$ has finitely many predecessors.
There are uncountably many orders on the square lattice
obeying these conditions (see e.g.~\cite{CR}).
Consider indeed a positive irrational number $\omega$,
and define the height of $(l,m)$ as $h_{l,m}=l+m\omega$.
The order obtained by defining $(l',m')<(l,m)$ as $h_{l',m'}<h_{l,m}$
has all required properties.
It is however not invariant under permutation of both axes.
One has indeed $(1,2)<(2,1)$ for $\omega<1$ and vice versa.
This construction easily extends to higher dimensions
by defining the height of a point as
$h_{l_1,\dots,l_D}=l_1\omega_1+\cdots+l_D\omega_D$,
where the numbers~$\omega_i$ are positive and linearly independent over the rationals.
A second fact is equally important.
Suppose we equip the square lattice with the total order associated with a given
irrational slope $\omega$.
The resulting records are just usual records,
albeit with lattice points being ordered according to increasing values of the height $h_{l,m}$.
As a consequence, the number $N$ of records among the first $L$ points
has the same statistics as in the case of usual, i.e., one-dimensional, records.
In brief,
both the novelty and the richness of the multidimensional record patterns studied here
are due to the fact that they are not based upon a total order, at variance with usual records.

The contents of the present paper is as follows.
In section~\ref{1d} we present a self-contained reminder on the statistics of usual records
and describe a backward recursive construction of those records for finite signals of length $L$.
The multidimensional record patterns obtained by extending the latter recursive construction
to higher-dimensional lattices are investigated in later sections.
Section~\ref{2d} is devoted to a detailed analysis of the two-dimensional case,
which is related to the fragmentation process studied in~\cite{spain,stick}.
We obtain exact combinatorial results on the statistics of the number~$N$
of records on finite rectangular samples of size $L\times M$,
as well as asymptotic results for large samples.
We also investigate the distribution of the landing position $K$ of the process,
i.e., the length of the stick adjacent to the origin.
The case of an arbitrary dimension~$D$ is addressed in section~\ref{hd}.
The main emphasis is on the asymptotic behavior of the full distribution
and of the moments of the number of record points on large hypercubic samples.
Section~\ref{disc} contains a brief summary of our main findings.
\ref{app} is devoted to the derivation of asymptotic large-$D$ expansions of various quantities.

\section{The one-dimensional case: usual records}
\label{1d}

\subsection{A reminder on record statistics}
\label{1dremind}

It is worth beginning with a self-contained reminder
on the statistics of usual records~\cite{gli,arn,nev,s+z,bun,wer,god}.
Consider a signal $x_n$ in discrete time ($n=1,2,\dots$),
modelled as a sequence of independent and identically distributed (iid)
real random variables with continuous distribution $\rho(x)$.
It is said that there is a record-breaking event at time~$n$,
or for short that time $n$ is a record,
if~$x_n$ is larger than all previous entries $x_{n'}$ ($n'=1,\dots,n-1$).
So, $n=1$ is always a record, whereas $n=2$ is a record with probability $1/2$, and so on.
It is indeed well established that there is a record at time $n$ with probability
\beq
p_n=\frac{1}{n},
\eeq
and that the occurrences of records at different times are statistically independent.

The key quantity of interest is the number of records,
$N=1,\dots,L$, on a finite signal of length~$L$.
The distribution $P_L(N)$ of this random number is conveniently encoded
in the generating function
\beq
G_L(z)=\sum_{N=1}^LP_L(N)z^N.
\label{gdef}
\eeq
The independence of the occurrences of records at different times yields the product formula
\beq
G_L(z)=\prod_{n=1}^L(1-p_n+zp_n)=\prod_{n=1}^L\frac{n-1+z}{n},
\eeq
i.e.,
\beq
G_L(z)=\frac{\Gamma(L+z)}{L!\,\Gamma(z)}=\frac{1}{L!}\sum_{N=1}^L\stir{L}{N}z^N,
\label{1dg}
\eeq
so that the distribution of $N$ reads
\beq
P_L(N)=\frac{1}{L!}\stir{L}{N},
\eeq
where the integers $\stir{L}{N}$,
referred to as the Stirling numbers of the first kind~\cite{stir},
are ubiquitous in combinatorics (see e.g.~\cite{knuth,GKP,FS}).

We introduce for further reference
the complementary cumulative distribu\-tion of the number of records $N$,
defined as the probability that the latter number exceeds a given $n$.
This quantity reads
\beq
F_L(n)=\prob{N>n}=\sum_{m=n+1}^LP_L(m)=\frac{1}{L!}\sum_{m=n+1}^L\stir{L}{m}.
\label{fdef}
\eeq

The mean value $\mean{N}_L$ and the variance $\cum{N^2}_L=\mean{N^2}_L-\mean{N}_L^2$
of the number of records read
\beqa
\mean{N}_L=\sum_{n=1}^Lp_n=H_L=\ln L+\gamma+\cdots,
\label{1dave}
\\
\cum{N^2}_L=\sum_{n=1}^Lp_n(1-p_n)=H_L-H_L^{(2)}=\ln L+\gamma-\frac{\pi^2}{6}+\cdots,
\label{1dvar}
\eeqa
with
\beq
H_L=\sum_{n=1}^L\frac{1}{n},\qquad
H_L^{(2)}=\sum_{n=1}^L\frac{1}{n^2},
\label{hdef}
\eeq
whereas $\gamma=0.577215\dots$ is Euler's constant,
and subleading terms go to zero.

The number of records $N$ takes its smallest and largest values with probabilities
\beq
P_L(1)=\frac{1}{L},\qquad
P_L(L)=\frac{1}{L!}.
\eeq

The distribution of the number of records exhibits a simple asymptotic behavior
for long signals ($L\gg1$).
All its cumulants indeed grow logarithmically with $L$,
with unit prefactor, i.e.,
\beq
\cum{N^k}_L=\ln L+a_k+\cdots,
\label{1dcum}
\eeq
where subleading terms go to zero.
To leading order, the distribution of $N$ thus becomes a Poissonian distribution
with parameter $\Lam=\ln L$.
The $a_k$ are numerical constants such that
\beq
\sum_{k\ge1}\frac{a_k}{k!}s^k=-\ln\Gamma(\e^s),
\eeq
i.e.,
\beqa
a_1=\gamma=0.577215\dots,
\nonumber\\
a_2=\gamma-\frac{\pi^2}{6}=-1.067718\dots,
\nonumber\\
a_3=\gamma-\frac{\pi^2}{2}+2\zeta(3)=-1.953472\dots,
\eeqa
and so on, where $\zeta$ denotes Riemann's zeta function.
The first two expressions agree with~(\ref{1dave}),~(\ref{1dvar}).

\subsection{Recursive construction}
\label{1drecur}

Records can be alternatively described by means of the following backward recursive construction.
In spite of its simplicity and of its efficiency,
the latter construction is not frequently met in the literature.
Let us however mention that it is the prototypical example
of a class of sample-space-reducing (SSR) processes put forward very recently~\cite{coro,yadav}.

Consider a finite signal of length $L$.

\begin{itemize}

\item
The largest record time $l_1$ is the time in the range $n=1,\dots,L$ for which $x_n$ is maximal.
It is therefore distributed uniformly in the range $l_1=1,\dots,L$.

\item
The second largest record time $l_2$ is the time in the range $n=1,\dots,l_1-1$
for which~$x_n$ is maximal.
It is therefore distributed uniformly in the range $l_2=1,\dots,l_1-1$, and so on.

\item
The process stops at step $N$ when $l_N=1$ is reached.

\end{itemize}

The number of records is in the range $N=1,\dots,L$.
In terms of the generating function $G_L(z)$ introduced in~(\ref{gdef}),
the first step of the construction translates to
\beq
G_L(z)=\frac{z}{L}\sum_{l=1}^LG_{l-1}(z),
\eeq
with initial condition $G_0(z)=1$.
The above equation is equivalent to the recursion
\beq
(L+1)G_{L+1}(z)=(L+z)G_L(z),
\label{1dgeq}
\eeq
whose solution coincides with~(\ref{1dg}).
This confirms that the above construction indeed yields the statistics of usual records.

\section{The two-dimensional case}
\label{2d}

\subsection{Recursive construction}
\label{2drecur}

Two-dimensional record patterns are defined by means of an extension
to the square lattice of the recursive construction described in section~\ref{1drecur}.
Consider a finite rectangular sample of size $L\times M$.
Lattice points,
to be referred to hereafter as record points, or records for short,
are deposited at random according to the following rules.

\begin{itemize}

\item
The first point $(l_1,m_1)$ is chosen
uniformly in the range $l_1=1,\dots,L$, $m_1=1,\dots,M$.

\item
The second point $(l_2,m_2)$ is chosen uniformly
in the range $l_2=1,\dots,l_1-1$, $m_2=1,\dots,m_1-1$, and so on.

\item
The process stops at step $N$,
when the $N$th point has either $l_N=1$ or $m_N=1$ or both.

\end{itemize}

The number of records is in the range
\beq
N=1,\dots,\min(L,M).
\eeq

The landing position of the process, defined as
\beq
K=\max(l_N,m_N)=1,\dots,\max(L,M),
\eeq
is nothing but the length of the stick adjacent to the origin
in the symmetric fragmentation process considered in~\cite{spain,stick}.

\begin{figure}[!ht]
\begin{center}
\includegraphics[angle=0,width=.6\linewidth,clip=true]{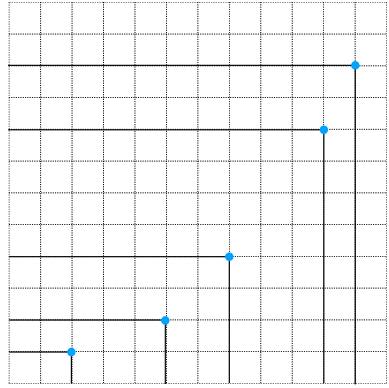}
\caption{\small
An instance of a two-dimensional record pattern on a square sample of size $L=M=12$,
with $N=5$ records at positions (11,10), (10,8), (7,4), (5,2) and (2,1),
and landing position $K=2$.}
\label{sample}
\end{center}
\end{figure}

This recursive construction is illustrated in figure~\ref{sample},
showing an instance of a two-dimensional record pattern on a square sample of size $L=M=12$,
with $N=5$ records at positions (11,10), (10,8), (7,4), (5,2) and (2,1),
and landing position $K=2$.

\subsection{Exact distribution of the number of records}
\label{2dstat}

The statistics of the number of records in two-dimensional record patterns
can be investigated by means of generating series.
This approach has been tackled in section~\ref{1drecur}.
Similar techniques were already used in~\cite{stick}.
Let
\beq
G_{L,M}(z)=\sum_NP_{L,M}(N)z^N
\eeq
be the generating function of the distribution
of the number of records on a finite sample of size $L\times M$.
The first step of the recursive construction described in section~\ref{2drecur} translates to
\beq
G_{L,M}(z)=\frac{z}{LM}\sum_{l=1}^L\sum_{m=1}^MG_{l-1,m-1}(z),
\eeq
with boundary conditions $G_{L,0}(z)=G_{0,M}(z)=1$.
The above equation is equivalent~to
\beqa
(L+1)(M+1)G_{L+1,M+1}(z)-(L+1)MG_{L+1,M}(z)
\nonumber\\
-L(M+1)G_{L,M+1}(z)+(LM-z)G_{L,M}(z)=0.
\label{gdiff}
\eeqa
This difference equation is recursive,
in the sense that it allows us to determine $G_{L,M}(z)$ step by step,
from the sole knowledge of boundary conditions.
However, at variance with~(\ref{1dgeq}), it cannot be solved by elementary means.
An integral representation of its solution can be derived as follows.
The triple generating function
\beq
G(x,y,z)=\sum_{L\ge0}\sum_{M\ge0}G_{L,M}(z)x^Ly^M
\eeq
obeys the partial differential equation
\beq
(1-x)(1-y)\frac{\partial^2G}{\partial x\partial y}=zG,
\label{gpde}
\eeq
with boundary conditions $G(x,0,z)=1/(1-x)$, $G(0,y,z)=1/(1-y)$.
Performing the change of variables
\beq
x=1-\e^{-\lam},\qquad y=1-\e^{-\mu},
\eeq
and setting $\G(\lam,\mu,z)=G(x,y,z)$, (\ref{gpde}) simplifies to
\beq
\frac{\partial^2\G}{\partial\lam\partial\mu}=z\G,
\eeq
with boundary conditions $\G(\lam,0,z)=\e^\lam$, $\G(0,\mu,z)=\e^\mu$.
The latter equation can be solved in Laplace space.
Defining for convenience the double Laplace transform as
\beq
\h\G(p,q,z)=\int_0^\infty\e^{-(p+1)\lam}\dd\lam\int_0^\infty\e^{-(q+1)\mu}\dd\mu\,\G(\lam,\mu,z),
\eeq
we obtain the rational expression
\beqa
\h\G(p,q,z)=R(p,q,z)=\frac{1}{pq}\frac{1-\eta(p,q)}{1-z\eta(p,q)},
\nonumber\\
\eta(p,q)=\frac{1}{(p+1)(q+1)}.
\eeqa
Performing the inverse transforms, we successively obtain
\beqa
&&\G(\lam,\mu,z)=
\int\frac{\dd p}{2\pi\ii}\e^{(p+1)\lam}
\int\frac{\dd q}{2\pi\ii}\e^{(q+1)\mu}
R(p,q,z),
\\
&&G(x,y,z)=
\int\frac{\dd p}{2\pi\ii}\frac{1}{(1-x)^{p+1}}
\int\frac{\dd q}{2\pi\ii}\frac{1}{(1-y)^{q+1}}
R(p,q,z),
\\
&&G_{L,M}(z)=
\int\frac{\dd p}{2\pi\ii}\frac{\Gamma(L+p+1)}{L!\,\Gamma(p+1)}
\int\frac{\dd q}{2\pi\ii}\frac{\Gamma(M+q+1)}{M!\,\Gamma(q+1)}
R(p,q,z),
\label{gint}
\eeqa
and finally
\beqa
P_{L,M}(N)
&=&
\int\frac{\dd p}{2\pi\ii}\frac{\Gamma(L+p+1)}{L!\,p\,\Gamma(p+1)}
\int\frac{\dd q}{2\pi\ii}\frac{\Gamma(M+q+1)}{M!\,q\,\Gamma(q+1)}
\nonumber\\
&\times&(1-\eta(p,q))\eta(p,q)^N.
\label{plmint}
\eeqa
In~(\ref{gint}) and~(\ref{plmint}),
integration contours in the complex $p$ and $q$-planes
circle once around all poles of the integrands.

The exact expression~(\ref{plmint}) for the distribution of the number
of two-dimensional records can be transformed by means of the identity
\beq
\int\frac{\dd p}{2\pi\ii}\frac{\Gamma(L+p+1)}{L!\,\Gamma(p+1)}\frac{1}{p(p+1)^{N+1}}=F_L(N),
\label{Fiden}
\eeq
which can be derived by applying residue calculus, using~(\ref{1dg}) and~(\ref{fdef}).
We thus obtain
\beq
P_{L,M}(N)=F_L(N-1)F_M(N-1)-F_L(N)F_M(N).
\label{plmff}
\eeq
The above formula has the following probabilistic interpretation.
The number $N$ of two-dimensional records on a rectangle of size $L\times M$
is distributed as the smaller of~$N_1$ and $N_2$,
which are respectively the numbers of one-dimensional records
on signals of sizes $L$ and $M$.
This property is a consequence of the recursive construction.
It is nevertheless comforting to see it emerging out of the formalism.

As a consequence of~(\ref{plmff}),
the smallest number of records ($N=1$) occurs with probability
\beq
P_{L,M}(1)=\frac{L+M-1}{LM},
\eeq
whereas the probability for the number of records to be the largest
only assumes a simple form for square samples:
\beq
P_{L,L}(L)=\frac{1}{L!^2}.
\eeq

\subsection{Mean number of records}
\label{2dave}

This section is devoted to a study of the mean number $\mean{N}_{L,M}$ of two-dimensional records,
with an emphasis on its asymptotic behavior on large square and rectangular samples.
We have
\beq
\mean{N}_{L,M}=\left(z\,\frac{\dd}{\dd z}\right)_{z=1}G_{L,M}(z).
\eeq
Equation~(\ref{gdiff}) yields the recursive difference equation
\beqa
(L+1)(M+1)\mean{N}_{L+1,M+1}-(L+1)M\mean{N}_{L+1,M}
\nonumber\\
-L(M+1)\mean{N}_{L,M+1}+(LM-1)\mean{N}_{L,M}=1,
\label{ndiff}
\eeqa
with boundary conditions $\mean{N}_{L,0}=\mean{N}_{0,M}=0$.
Exact rational expressions for the mean number of records can be obtained
by solving the above equation recursively.
Table~\ref{n12tab} shows a comparison between $\mean{N}_L$ for one-dimensional signals of length $L$
and $\mean{N}_{L,L}$ for two-dimensional square samples of size $L\times L$.
The complexity of mean number of records grows much faster in two dimensions than in one dimension.

\begin{table}[!ht]
\begin{center}
\begin{tabular}{|l|c|c|c|c|c|c|c|c|c|}
\hline
$L$ & 1 & 2 & 3 & 4 & 5 & 6 & 7 & 8 & 9\cr
\hline
$\mean{N}_L$ &
1 & $\frac{3}{2}$ & $\frac{11}{6}$ & $\frac{25}{12}$ & $\frac{137}{60}$ &
$\frac{49}{20}$ & $\frac{363}{140}$ & $\frac{761}{280}$ & $\frac{7\,129}{2\,520}$\cr
\hline
$\mean{N}_{L,L}$ &
1 & $\frac{5}{4}$ & $\frac{53}{36}$ & $\frac{475}{288}$ & $\frac{4\,309}{2\,400}$ &
$\frac{497\,567}{259\,200}$ & $\frac{25\,752\,527}{12\,700\,800}$ &
$\frac{862\,916\,227}{406\,425\,600}$ & $\frac{24\,239\,088\,877}{10\,973\,491\,200}$\cr
\hline
\end{tabular}
\caption
{Exact rational expressions for the mean number $\mean{N}_{L}$ of records
for one-dimensional signals of length $L$
and $\mean{N}_{L,L}$ for two-dimensional square samples of size $L\times L$, up to $L=9$.}
\label{n12tab}
\end{center}
\end{table}

Equation~(\ref{gint}) yields the following exact integral representation for the mean number of records:
\beqa
\mean{N}_{L,M}&=&
\int\frac{\dd p}{2\pi\ii}\frac{\Gamma(L+p+1)}{L!\,\Gamma(p+1)}
\int\frac{\dd q}{2\pi\ii}\frac{\Gamma(M+q+1)}{M!\,\Gamma(q+1)}
\nonumber\\
&\times&\frac{1}{pq(p+q+pq)},
\label{nlmint}
\eeqa
which is suitable for studying the behavior of $\mean{N}_{L,M}$ on large samples.
Neglecting terms of relative order $1/L$ and $1/M$,
the above expression simplifies to
\beq
\mean{N}_{L,M}\approx
\int\frac{\dd p}{2\pi\ii}\frac{L^p}{\Gamma(p+1)}
\int\frac{\dd q}{2\pi\ii}\frac{M^q}{\Gamma(q+1)}
\frac{1}{pq(p+q+pq)}.
\label{n2pq}
\eeq

Several kinds of sample shapes can be considered.
In the simplest situation of large square samples ($L=M\gg1$),
performing the integral over $q$ in~(\ref{n2pq}) yields
\beq
\mean{N}_{L,L}\approx\Lam+\gamma
-\int\frac{\dd p}{2\pi\ii}\frac{\e^{\Lam p^2/(p+1)}}{p^2\Gamma(p+1)\Gamma(1/(p+1))},
\label{n2p}
\eeq
with $\Lam=\ln L$ and $\gamma$ is again Euler's constant.
For large $\Lam$, the integral is dominated by small values of $p$
such that $\Lam p^2$ is of order unity.
To leading order, this reads
\beq
\int\frac{\dd p}{2\pi\ii}\frac{\e^{\Lam p^2}}{p^2}=\sqrt\frac{\Lam}{\pi}.
\label{leadint}
\eeq
Higher-order corrections can be derived systematically by expanding the integrand in~(\ref{n2p})
in increasing powers of $p$ and integrating term by term.
We thus obtain an asymptotic expansion in decreasing powers of $\ln L$:
\beq
\mean{N}_{L,L}=\ln L-\sqrt\frac{\ln L}{\pi}+\gamma
-\frac{1}{2\sqrt{\pi\ln L}}\left(\gamma-\frac{1}{8}-\frac{\pi^2}{6}\right)+\cdots
\label{n2sq}
\eeq

For large rectangular samples,
i.e., for $L$ and $M$ large and proportional to each other,
performing the integral over $q$ in~(\ref{n2pq}) yields
\beq
\mean{N}_{L,M}\approx\Lam-\frac{r}{2}+\gamma
-\int\frac{\dd p}{2\pi\ii}\frac{\e^{\Lam p^2/(p+1)-rp(p+2)/(2(p+1))}}
{p^2\Gamma(p+1)\Gamma(1/(p+1))},
\label{n2int}
\eeq
with
\beq
\Lam=\frac{\ln LM}{2},\qquad r=\ln\frac{M}{L}.
\eeq
In the regime where $\Lam$ is large whereas $r$ remains finite,
the integral is again dominated by small values of $p$ such that $\Lam p^2$ is of order unity.
We are thus left with
\beqa
\mean{N}_{L,M}&\approx&\frac{\ln LM}{2}-\sqrt\frac{\ln LM}{2\pi}+\gamma
\nonumber\\
&-&\frac{1}{\sqrt{2\pi\ln LM}}
\left[\frac{1}{2}\left(\ln\frac{M}{L}\right)^2+\gamma-\frac{1}{8}-\frac{\pi^2}{6}\right]+\cdots
\label{n2rect}
\eeqa

For very anisotropic samples such that $1\ll L\ll M$,
the integral over $q$ in~(\ref{n2pq}) is dominated by the trivial pole at $q=0$,
and so the two-dimensional result boils down to the one-dimensional one:
\beq
\mean{N}_{L,M}\approx{N}_L\approx\ln L+\gamma,
\label{n2anis}
\eeq
irrespective of the larger side $M$.
This reduction is a manifestation of the property expressed by~(\ref{plmff}).
We recall that the distributions of $N_1$ and $N_2$ are nearly Poissonian,
and therefore sharply peaked around $\Lam_1=\ln L$ and $\Lam_2=\ln M$.
If $\Lam_2$ is sufficiently larger than $\Lam_1$,
the distributions of $N_1$ and $N_2$ have essentially no overlap,
and the two-dimensional number of records $N$ is essentially distributed as $N_1$.

The crossover between the results~(\ref{n2sq}) in the isotropic geometry of squares
and~(\ref{n2anis}) in the very anisotropic one
takes place when $r=\Lam_2-\Lam_1$ is positive and of the order of $\sqrt\Lam$.
This regime can be investigated by simplifying the integral representation~(\ref{n2int}) as
\beq
\mean{N}_{L,M}\approx\Lam-\frac{r}{2}+\gamma
-\int\frac{\dd p}{2\pi\ii}\frac{\e^{\Lam p^2-rp}}{p^2},
\eeq
yielding
\beq
\mean{N}_{L,M}\approx\Lam+\gamma
-\sqrt\frac{\Lam}{\pi}\,\e^{-r^2/(4\Lam)}
-\frac{r}{2}\erf\frac{r}{2\sqrt{\Lam}},
\eeq
where $\erf$ denotes the error function.
In the weakly anisotropic regime ($r\ll\sqrt\Lam$), the expansion
\beq
\mean{N}_{L,M}\approx\Lam-\sqrt\frac{\Lam}{\pi}+\gamma-\frac{r^2}{4\sqrt{\pi\Lam}}+\cdots
\eeq
matches~(\ref{n2rect}).
In the strongly anisotropic regime ($r\gg\sqrt\Lam$), we have
\beq
\mean{N}_{L,M}\approx\ln L+\gamma-2\sqrt\frac{\Lam^3}{\pi}\frac{\e^{-r^2/(4\Lam)}}{r^2}.
\eeq
The correction to~(\ref{n2anis}) is exponentially small, and negative,
in agreement with the property expressed by~(\ref{plmff}).

\subsection{Second moment and variance of the number of records}
\label{2dvar}

Higher moments of the number of two-dimensional records can be dealt with along the same lines.
The second moment is given by
\beq
\mean{N^2}_{L,M}=\left(z\,\frac{\dd}{\dd z}\right)^2_{z=1}G_{L,M}(z).
\eeq
Equation~(\ref{gdiff}) yields the recursive difference equation
\beqa
(L+1)(M+1)\mean{N^2}_{L+1,M+1}-(L+1)M\mean{N^2}_{L+1,M}
\nonumber\\
-L(M+1)\mean{N^2}_{L,M+1}+(LM-1)\mean{N^2}_{L,M}=2\mean{N}_{L,M}+1,
\label{n2diff}
\eeqa
with boundary conditions $\mean{N^2}_{L,0}=\mean{N^2}_{0,M}=0$.
Exact rational expressions for $\mean{N^2}_{L,M}$ can be obtained
by solving~(\ref{ndiff}) and~(\ref{n2diff}) step by step.
Table~\ref{v12tab} shows a comparison between the variances of the number of records
$\cum{N^2}_L$ for one-dimensional signals of length $L$
and $\cum{N^2}_{L,L}$ for two-dimensional square samples of size $L\times L$.
The complexity of these exact results again grows much faster in two dimensions than in one dimension.

\begin{table}[!ht]
\begin{center}
\begin{tabular}{|l|c|c|c|c|c|c|c|c|c|}
\hline
$L$ & 1 & 2 & 3 & 4 & 5 & 6\cr
\hline
$\cum{N^2}_L$ &
0 & $\frac{1}{4}$ & $\frac{17}{36}$ & $\frac{95}{144}$ & $\frac{2\,951}{3\,600}$ &
$\frac{3\,451}{3\,600}$\cr
\hline
$\cum{N^2}_{L,L}$ &
0 & $\frac{3}{16}$ & $\frac{395}{1\,296}$ & $\frac{33\,575}{82\,944}$ & $\frac{2\,826\,119}{5\,760\,000}$ &
$\frac{38\,000\,939\,711}{67\,184\,640\,000}$\cr
\hline
\end{tabular}
\caption
{Exact rational expressions for the variances of the number of records
$\cum{N^2}_L$ for one-dimensional signals of length $L$
and $\cum{N^2}_{L,L}$ for two-dimensional square samples of size $L\times L$, up to $L=6$.}
\label{v12tab}
\end{center}
\end{table}

Equation~(\ref{gint}) yields the exact integral representation
\beqa
\mean{N^2}_{L,M}&=&
\int\frac{\dd p}{2\pi\ii}\frac{\Gamma(L+p+1)}{L!\,\Gamma(p+1)}
\int\frac{\dd q}{2\pi\ii}\frac{\Gamma(M+q+1)}{M!\,\Gamma(q+1)}
\nonumber\\
&\times&\frac{2+p+q+pq}{pq(p+q+pq)^2},
\eeqa
which is again very suitable for studying the behavior
of $\mean{N^2}_{L,M}$ on large samples.
Neglecting terms of relative order $1/L$ and $1/M$,
the above expression simplifies to
\beq
\mean{N^2}_{L,M}\approx
\int\frac{\dd p}{2\pi\ii}\frac{L^p}{\Gamma(p+1)}
\int\frac{\dd q}{2\pi\ii}\frac{M^q}{\Gamma(q+1)}
\frac{2+p+q+pq}{pq(p+q+pq)^2}.
\label{n22pq}
\eeq

Let us focus our attention onto the simplest situation of large square samples ($L=M\gg1$).
Performing the integral over $q$ in~(\ref{n22pq}) yields
\beqa
\mean{N^2}_{L,L}&\approx&(\Lam+\gamma)^2+\frac{\pi^2}{6}
-\int\frac{\dd p}{2\pi\ii}\frac{\e^{\Lam p^2/(p+1)}}{\Gamma(p+1)\Gamma(1/(p+1))}
\nonumber\\
&\times&\left\{\frac{2}{p^3}+\frac{1}{p^2}+
\frac{2}{p^2(p+1)}\left[\Lam-\psi
\left(\frac{1}{p+1}\right)
\right]\right\},
\eeqa
where $\Lam=\ln L$
and $\psi(z)=\Gamma'(z)/\Gamma(z)$ is the digamma function.
For large $\Lam$, the integral is again dominated by small values of $p$
such that $\Lam p^2$ is of order unity.
The same analysis as in section~\ref{2dave} yields the following expansion
for the variance of the number of records in decreasing powers of $\ln L$:
\beqa
\cum{N^2}_{L,L}&=&\left(1-\frac{1}{\pi}\right)\left(\ln L+\gamma-\frac{\pi^2}{6}\right)
-\sqrt\frac{\ln L}{4\pi}+\frac{1}{8\pi}
\nonumber\\
&-&\frac{1}{4\sqrt{\pi\ln L}}\left(\gamma+\frac{1}{8}-\frac{5\pi^2}{6}+4\zeta(3)\right)+\cdots
\label{nvarsq}
\eeqa

\begin{figure}[!ht]
\begin{center}
\includegraphics[angle=0,width=.6\linewidth]{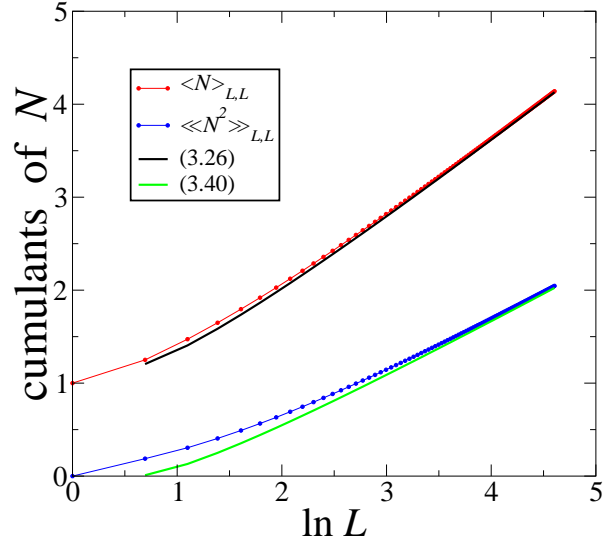}
\caption{\small
First two cumulants of the number of records $N$ on square samples,
against $\ln L$.
Exact values for finite sample sizes up to $L=100$ are compared to
the four-term asymptotic expansions~(\ref{n2sq}) and~(\ref{nvarsq}).}
\label{2dplot}
\end{center}
\end{figure}

To sum up,
the main characteristic features of the statistics of two-dimensional record patterns are as follows.
The number of records $N$ on a rectangular sample of size $L\times M$ is distributed as the
smaller of the numbers $N_1$ and $N_2$ of one-dimensional records
for sequences of lengths $L$ and $M$ (see~(\ref{plmff})).
For large square samples,
the mean number of records $\mean{N}_{L,L}$ exhibits a negative correction in $\sqrt{\ln L/\pi}$
with respect to the one-dimensional case of usual records (see~(\ref{n2sq})),
whereas the logarithmic growth of its variance
is reduced by the universal factor $1-1/\pi$ (see~(\ref{nvarsq})).
These results are illustrated in figure~\ref{2dplot},
showing the first two cumulants of the number of records $N$ on square samples against $\ln L$.
The exact values of these quantities for finite sample sizes up to $L=100$,
obtained by solving the recursive difference equations~(\ref{ndiff}) and~(\ref{n2diff}),
are accurately described by the asymptotic expansions~(\ref{n2sq}) and~(\ref{nvarsq}).

In section~\ref{hd}
we extend the above results to higher-dimensional record patterns in any dimension $D$.

\subsection{Distribution of the landing position}
\label{2dk}

The formalism used in section~\ref{2dstat} to derive the exact distribution of the number
of records can be extended to investigate other quantities of interest
in two-dimensional record patterns.

Hereafter we consider the distribution $R_{L,M}(K)$ of the landing position $K$,
defined in section~\ref{2drecur} and illustrated in figure~\ref{sample}.
This quantity equals the length of the stick adjacent to the origin
in the symmetric fragmentation process considered in~\cite{spain,stick}.

The first step of the recursive construction described in section~\ref{2drecur} translates to
\beq
R_{L,M}(K)=\frac{1}{LM}\sum_{l=1}^L\sum_{m=1}^MR_{l-1,m-1}(K),
\label{ksum}
\eeq
with boundary conditions $R_{L,0}(K)=\delta_{L,K-1}$,
$R_{0,M}(K)=\delta_{M,K-1}$, where $\delta_{i,j}$ denotes the Kronecker symbol.
Indeed,
if $l_1=1,\dots,L$ and $m_1=1$ are chosen at the first step,
the process stops at this step and the landing position is $K=l_1$.
Similarly,
if $l_1=1$ and $m_1=1,\dots,M$ are chosen at the first step,
the process stops at this step and the landing position is $K=m_1$.
Equation~(\ref{ksum}) is equivalent to
\beqa
(L+1)(M+1)R_{L+1,M+1}(K)-(L+1)MR_{L+1,M}(K)
\nonumber\\
-L(M+1)R_{L,M+1}(K)+(LM-1)R_{L,M}(K)=0.
\label{krdiff}
\eeqa
The moments of the landing position,
\beq
\mean{K^s}_{L,M}=\sum_K K^s\,R_{L,M}(K),
\label{kmomdef}
\eeq
obey the same homoge\-neous difference equation, i.e.,
\beqa
(L+1)(M+1)\mean{K^s}_{L+1,M+1}-(L+1)M\mean{K^s}_{L+1,M}
\nonumber\\
-L(M+1)\mean{K^s}_{L,M+1}+(LM-1)\mean{K^s}_{L,M}=0,
\label{kmomdiff}
\eeqa
with boundary conditions $\mean{K^s}_{L,0}=(L+1)^s$, $\mean{K^s}_{0,M}=(M+1)^s$.
The exponent $s$ is not necessarily an integer.

The difference equations~(\ref{krdiff}) and~(\ref{kmomdiff}) are recursive,
just as~(\ref{gdiff}),~(\ref{ndiff}) and~(\ref{n2diff}).
All these equations can be solved recursively, from the sole knowledge of boundary conditions.
As an illustration of the method,
table~\ref{ktab} gives the exact rational values of the first two moments of the landing position $K$
for square samples of size~$L\times L$.
The asymptotic growth of these moments is given in~(\ref{mom1st}).

\begin{table}[!ht]
\begin{center}
\begin{tabular}{|l|c|c|c|c|c|c|c|c|}
\hline
$L$ & 1 & 2 & 3 & 4 & 5 & 6 & 7 & 8\cr
\hline
$\mean{K}_{L,L}$ &
1 & $\frac{3}{2}$ & $\frac{11}{6}$ & $\frac{203}{96}$ & $\frac{227}{96}$ &
$\frac{22\,399}{8\,640}$ & $\frac{211\,943}{75\,600}$ & $\frac{38\,719\,637}{12\,902\,400}$\cr
\hline
$\mean{K^2}_{L,L}$ &
1 & $\frac{5}{2}$ & $\frac{71}{18}$ & $\frac{1\,583}{288}$ & $\frac{10\,307}{1\,440}$ &
$\frac{15\,413}{1\,728}$ & $\frac{814\,511}{75\,600}$ & $\frac{492\,106\,337}{38\,707\,200}$\cr
\hline
\end{tabular}
\caption
{Exact rational expressions for the first two moments of the landing position $K$
for square samples of size $L\times L$, up to $L=8$.}
\label{ktab}
\end{center}
\end{table}

An integral representation of $R_{L,M}(K)$ can be derived by means
of the generating-function approach presented in section~(\ref{2dstat}).
We thus obtain
\beqa
R_{L,M}(K)&=&
\int\frac{\dd p}{2\pi\ii}\frac{\Gamma(L+p+1)}{L!\,\Gamma(p+1)}
\int\frac{\dd q}{2\pi\ii}\frac{\Gamma(M+q+1)}{M!\,\Gamma(q+1)}
\nonumber\\
&\times&\frac{S(p,q,K)}{p+q+pq},
\label{rlmint}
\eeqa
with
\beq
S(p,q,K)=(K-1)!\!\left(\frac{\Gamma(p+2)}{\Gamma(K+p+1)}+\frac{\Gamma(q+2)}{\Gamma(K+q+1)}
-\delta_{K,1}\right)\!.
\eeq

The above expression is again suitable for studying the regime of large samples.
Neglecting terms of relative order $1/L$ and $1/M$,~(\ref{rlmint}) simplifies to
\beq
R_{L,M}(K)\approx
\int\frac{\dd p}{2\pi\ii}\frac{L^p}{\Gamma(p+1)}
\int\frac{\dd q}{2\pi\ii}\frac{M^q}{\Gamma(q+1)}
\frac{S(p,q,K)}{p+q+pq}.
\label{rasyint}
\eeq
Henceforth we again focus our attention onto the simplest situation of large square samples ($L=M\gg1$).
Rearranging terms in $S(p,q,K)$ and performing the integral over~$q$ in~(\ref{rasyint}) yields
\beq
R_{L,L}(K)\approx
2\left(1-\half\delta_{K,1}\right)
\int\frac{\dd p}{2\pi\ii}\frac{\e^{\Lam p^2/(p+1)}}{\Gamma(1/(p+1))}\frac{(K-1)!}{\Gamma(K+p+1)},
\label{rllint}
\eeq
with $\Lam=\ln L$.

Let us first consider the asymptotic behavior of the distribution $R_{L,L}(K)$
in the regime where the sample size $L$ becomes large, whereas the landing position $K$ is finite.
The integral in~(\ref{rllint}) is again dominated by small values of $p$
such that~$\Lam p^2$ is of order unity.
We thus obtain, to leading order
\beq
R_{L,L}(K)\approx\frac{1}{\sqrt{\pi\ln L}}\frac{1-\half\delta_{K,1}}{K}.
\label{rklaw}
\eeq
This result calls for several remarks.
The probability of having $K=1$ is reduced by a factor 2 with respect to the rest of the distribution.
This property holds for the full distribution of generic sticks as well~\cite{stick}.
Here, $K=1$ occurs only when the last record is deposited at $(1,1)$,
whereas $K\ge2$ corresponds to two distinct positions $(1,K)$ and $(K,1)$ for the last record,
i.e., two distinct orientations for the stick adjacent to the origin.
The $1/K$ law entering~(\ref{rklaw}) is not normalizable if extended to arbitrarily large values of $K$.
The dependence of the prefactor on the sample size $L$
suggests that the above law is cutoff at the scale $\ln K\sim\sqrt{\ln L}$.
This estimate is corroborated by the scaling result~(\ref{rllsca}).

Higher-order corrections to~(\ref{rklaw}) can be derived by expanding the integrand in~(\ref{rllint})
in increasing powers of $p$ and integrating term by term.
The first non-trivial correction reads
\beq
R_{L,L}(K)=\frac{1}{\sqrt{\pi\ln L}}\frac{1-\half\delta_{K,1}}{K}\left(1-\frac{C(K)}{4\ln L}+\cdots\right),
\label{rll1}
\eeq
with
\beqa
C(K)&=&(H_K)^2-3H_K+H_K^{(2)}-\frac{\pi^2}{3}+2\gamma+\frac{3}{4}
\nonumber\\
&=&(\ln K)^2+(2\gamma-3)\ln K-\frac{\pi^2}{6}+\gamma^2-\gamma+\frac{3}{4}+\cdots,
\label{ckres}
\eeqa
where $H_K$ and $H_K^{(2)}$ are defined in~(\ref{hdef}).

We now turn to the regime where the sample size $L$ and the landing position $K$ are both large.
The relevant scaling corresponds to $\Lam=\ln L$ and $\lambda=\ln K$ being large and comparable,
with a fixed ratio $a$, so that $\lambda=a\Lam$, i.e.,
\beq
K=L^a\qquad(0<a<1).
\eeq
The formula~(\ref{rllint}) then simplifies to
\beq
R_{L,L}(K=L^a)\approx
2\int\frac{\dd p}{2\pi\ii}\frac{\e^{\Lam(p^2/(p+1)-(p+1)a)}}{\Gamma(1/(p+1))}.
\label{rllintsca}
\eeq
It is legitimate to estimate the above integral by means of the saddle-point method.
The saddle point reads $p=-1+1/\sqrt{1-a}$.
We thus obtain
\beq
R_{L,L}(K=L^a)\approx
\frac{L^{-2(1-\sqrt{1-a})}}{\sqrt{\pi\ln L}\,(1-a)^{3/4}\,\Gamma(\sqrt{1-a})}.
\label{rllres}
\eeq

The most probable values of the landing position correspond to $a\ll1$,
where the result~(\ref{rllres}) simplifies to
\beq
R_{L,L}(K)\approx\frac{1}{\sqrt{\pi\ln L}}\frac{\e^{-(\ln K)^2/(4\ln L)}}{K}.
\label{rllsca}
\eeq
We therefore predict that the bulk of the distribution of $K$ assumes a half-log-normal form.
This scaling form matches to leading order the growth of the amplitude~$C(K)$ (see~(\ref{ckres})).

An alternative viewpoint consists in considering the moments of the landing position,
introduced in~(\ref{kmomdef}), for $s>0$.
The most direct route to determine the growth of these moments for large $L$ starts from~(\ref{rllintsca})
and yields
\beq
\mean{K^s}_{L,L}\approx
2\int\frac{\dd p}{2\pi\ii}\frac{\e^{\Lam(p^2/(p+1))}}{\Gamma(1/(p+1))}\frac{1}{p-s},
\eeq
where the contour integral is to the right of the pole at $p=s$.
The integral is dominated by the latter pole.
We thus obtain the scaling law
\beq
\mean{K^s}_{L,L}\approx\frac{2}{\Gamma(1/(s+1))}L^{s^2/(s+1)}.
\label{momsca}
\eeq
In particular, integer moments of $K$ grow with rational exponents, i.e.,
\beqa
&&\mean{K}_{L,L}\approx\frac{2}{\sqrt\pi}\,L^{1/2},
\nonumber\\
&&\mean{K^2}_{L,L}\approx\frac{2}{\Gamma(1/3)}\,L^{4/3},
\nonumber\\
&&\mean{K^3}_{L,L}\approx\frac{2}{\Gamma(1/4)}\,L^{9/4},
\label{mom1st}
\eeqa
and so on.

The results~(\ref{rllres}) and~(\ref{momsca}) demonstrate multifractality~\cite{paladin,stanley}.
The probability distribution of the landing position and its moments indeed scale as
\beq
R_{L,L}(K=L^a)\sim L^{-\phi(a)},\qquad
\mean{K^s}_{L,L}\sim L^{\tau(s)}.
\eeq
Both nonlinear spectra of exponents
\beqa
\phi(a)&=&2(1-\sqrt{1-a})\qquad(0<a<1),
\nonumber\\
\tau(s)&=&\frac{s^2}{s+1}{\hskip 61.5pt}(s>0)
\eeqa
encode the multifractality of the distribution of the landing position.
They are related to each other by the Legendre transformation
\beq
\tau(s)=\max_a\left[(s+1)a-\phi(a)\right].
\eeq

We note that moments with a negative exponent ($s=-\sigma<0$) escape multi\-fractality.
They are dominated by microscopic values of $K$, where~(\ref{rklaw}) holds, hence
\beq
\mean{K^{-\sigma}}_{L,L}\approx\frac{1}{\sqrt{\pi\ln L}}\left(\zeta(\sigma+1)-\frac{1}{2}\right).
\eeq

\section{The higher-dimensional case}
\label{hd}

\subsection{Recursive construction}
\label{hdrecur}

Higher-dimensional record patterns are defined by means of an extension
of the recursive construction described in sections~\ref{1drecur} and~\ref{2drecur}.
Consider a finite parallelotopic sample of size $L_1\times\dots\times L_D$,
with side lengths $L_i\ge1$ for $i=1,\dots,D$.
Records are deposited onto this sample according to the following rules.

\begin{itemize}

\item
The first point $(l_{1,1},\dots,l_{1,D})$ is chosen
uniformly in the range $l_{1,i}=1,\dots,L_i$ for $i=1,\dots,D$.

\item
The second point $(l_{2,1},\dots,l_{2,D})$ is chosen
uniformly in the range $l_{2,i}=1,\dots,l_{1,i}-1$ for $i=1,\dots,D$, and so on.

\item
The process stops at step $N$,
when (at least) one of the co-ordinates $l_{N,i}$ of the $N$th point equals unity.

\end{itemize}

The number of records is in the range $N=1,\dots,\min\left(\li\right)$.

\subsection{Exact distribution of the number of records}
\label{hdstat}

The statistics of the number $N$ of $D$-dimensional records
in the random patterns thus generated can be investigated along the lines of the approach
sketched for $D=1$ in section~\ref{1drecur} and described for $D=2$ in section~\ref{2dstat}.
We shall henceforth only give essential results.

The distribution of $N$ is exactly given by the following formula,
which is a direct generalization of~(\ref{plmint}):
\beqa
P_{\li}(N)&=&\int\etc\int
\prod_{i=1}^D\frac{\dd p_i}{2\pi\ii}\frac{\Gamma(L_i+p_i+1)}{L_i!\,p_i\,\Gamma(p_i+1)}
\nonumber\\
&\times&
\left(1-\eta\left(\pei\right)\right)\eta\left(\pei\right)^N,
\label{pliint}
\eeqa
with
\beq
\eta\left(\pei\right)=\prod_{i=1}^D\frac{1}{1+p_i}.
\eeq

The above expression can be transformed by means of the identity~(\ref{Fiden}) into
\beq
P_{\li}(N)=\prod_{i=1}^DF_{L_i}(N-1)-\prod_{i=1}^DF_{L_i}(N),
\label{pliff}
\eeq
whose meaning is that the number $N$ of records
is distributed as the smallest of the numbers $N_i$ of one-dimensional records
of independent signals whose lengths $L_i$ are the sides of the sample.
In particular, the probability of having one single record reads
\beq
P_{\li}(1)=1-\prod_{i=1}^D\left(1-\frac{1}{L_i}\right).
\eeq

\subsection{Asymptotic behavior on large hypercubic samples}
\label{hdcubic}

We now turn to the asymptotic behavior of the distribution of the number $N$ of records,
and especially of its mean value and variance.
For definiteness we restrict the analysis to large hypercubic samples,
with equal sides in all directions ($L_i=L$ for $i=1,\dots,D$).
The formula~(\ref{pliff}) simplifies to
\beq
P_{\lll}(N)=F_L(N-1)^D-F_L(N)^D,
\label{pllff}
\eeq
meaning that the number $N$ of $D$-dimensional records
is distributed as the smallest of $D$ independent copies
of the number $N_1$ of one-dimensional records in a signal of length~$L$.

The result~(\ref{pllff}) is the starting point of the following asymptotic analysis,
which is entirely based on probabilistic reasoning.
Let us first revisit the one-dimensional case of usual records.
To leading order as $\Lam=\ln L$ is large,
the bulk of the distribution of $N_1$ is Gaussian,
and its mean value and its variance are both
equal to $\Lam$ (see~(\ref{1dave}) and~(\ref{1dvar})).
We have therefore
\beq
N_1\approx\ln L-\sqrt{\ln L}\,X_1,
\eeq
where the rescaled variable $X_1$ is a standard Gaussian variable,
with $\mean{X_1}=0$ and $\mean{X_1^2}=1$.
The minus sign has been introduced for further convenience.
The corresponding probability density and cumulative distribution read
\beqa
f(X)=\Phi'(X)=\frac{\e^{-X^2/2}}{\sqrt{2\pi}},
\label{fphi1}
\\
\Phi(x)=\prob{X_1<x}
=\int_{-\infty}^xf(X)\,\dd X=\frac{1}{2}\left(1+\erf\frac{x}{\sqrt{2}}\right).
\label{fphi2}
\eeqa

\begin{figure}[!ht]
\begin{center}
\includegraphics[angle=0,width=.6\linewidth]{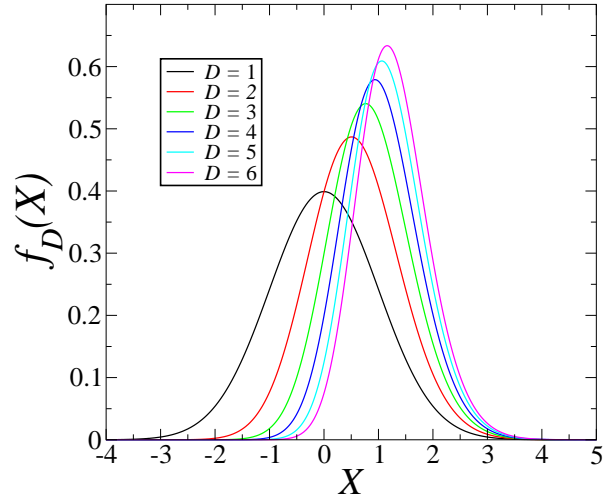}
\caption{\small
Probability density $f_D(X)$ of the rescaled variable $X$
descri\-bing the fluctuations of the number of $D$-dimensional records on large hypercubic samples
(see~(\ref{xdef}) and~(\ref{fd})) for several dimensions $D$ (see legend).}
\label{fplot}
\end{center}
\end{figure}

For $D$-dimensional records on large hypercubic samples,
as a consequence of~(\ref{pllff}),
we have a similar scaling law, i.e.,
\beq
N\approx\ln L-\sqrt{\ln L}\,X,
\label{xdef}
\eeq
where the rescaled variable $X$ is distributed as the largest of $D$ standard Gaussian variables.
In other words, the cumulative distribution of $X$ reads
\beq
\prob{X<x}=\Phi(x)^D.
\label{phid}
\eeq
Its probability density is therefore given by
\beq
f_D(X)=D\Phi(X)^{D-1}f(X)
=\frac{D}{\sqrt{2\pi}}\Phi(X)^{D-1}\e^{-X^2/2}.
\label{fd}
\eeq
Figure~\ref{fplot} demonstrates that the probability density $f_D(X)$
slowly drifts towards positive values while its width shrinks as $D$ increases.
The black curve shows the Gaussian density $f_1(X)=f(X)$ (see~(\ref{fphi1})).

There is therefore a direct relationship between multidimensional records
and extreme-value statistics for Gaussian random variables.
In particular,
the mean and variance of the number of records $N$ on large hypercubic samples read
\beqa
&&\mean{N}_\lll\approx\ln L-A_D\sqrt{\ln L},
\nonumber\\
&&\cum{N^2}_\lll\approx B_D\ln L,
\label{asymoms}
\eeqa
where the amplitudes $A_D$ and $B_D$ are given by
\beqa
A_D&=&\mean{X}_D=\int_{-\infty}^\infty X\,f_D(X)\,\dd X,
\nonumber\\
C_D&=&\mean{X^2}_D=\int_{-\infty}^\infty X^2\,f_D(X)\,\dd X,
\nonumber\\
B_D&=&\cum{X^2}_D=C_D-A_D^2,
\label{abcmoms}
\eeqa
where $f_D(X)$ is the probability density
of the largest of~$D$ standard Gaussian variables (see~(\ref{fd})).

The above integrals can be evaluated in closed form up to $D=5$
by means of an elegant approach due to Selby~\cite{note}.
First, integrations by parts are used to bring $A_D$ and $C_D$ to the form
\beqa
A_D=\frac{D(D-1)}{2\pi}\int_{-\infty}^\infty\e^{-X^2}\Phi(X)^{D-2}\,\dd X,
\nonumber\\
C_D=1+\frac{D(D-1)(D-2)}{2(2\pi)^{3/2}}\int_{-\infty}^\infty\e^{-3X^2/2}\Phi(X)^{D-3}\,\dd X.
\label{acint}
\eeqa
Second, noticing that
\beq
E(X)=\Phi(X)-\frac{1}{2}=\frac{1}{2}\erf\frac{X}{\sqrt{2}}
\eeq
is an odd function of $X$,
the integrals entering~(\ref{acint}) can be expanded as
\beqa
A_D=\frac{D(D-1)}{2^{D-1}\pi}
\sum_{n=0}^{\Int{(D-2)/2}}2^{2n}{D-2\choose2n}I_n(2),
\nonumber\\
C_D=1+\frac{D(D-1)(D-2)}{2^{D-2}(2\pi)^{3/2}}
\sum_{n=0}^{\Int{(D-3)/2}}2^{2n}{D-3\choose2n}I_n(3),
\eeqa
with
\beq
I_n(s)=\int_{-\infty}^\infty\e^{-sX^2/2}E(X)^{2n}\,\dd X.
\eeq
Third, the key observation that the $I_n(s)$ obey
`a strange reduction rule'~\cite{note}, i.e.,
\beq
I_{n+1}(s)=\frac{2n+1}{2\pi\sqrt{s}}\int_0^{1/\sqrt{s}}\frac{\dd y}{1+y^2}\,I_n(1+s(1+y^2)),
\eeq
allows one to derive
\beq
I_0(s)=\sqrt\frac{2\pi}{s},\qquad
I_1(s)=\frac{1}{\sqrt{2\pi s}}\arcsin\frac{1}{1+s}.
\eeq
The higher integrals $I_n(s)$ for $n\ge2$ do not seem to admit closed-form expressions.
This approach yields
\beqa
A_1&=&0,
\nonumber\\
A_2&=&\frac{1}{\sqrt{\pi}}=0.564189\dots,
\nonumber\\
A_3&=&\frac{3}{2\sqrt{\pi}}=0.846284\dots,
\nonumber\\
A_4&=&\frac{3\arccos(-1/3)}{\pi^{3/2}}=1.029375\dots,
\nonumber\\
A_5&=&\frac{5\arccos(-23/27)}{2\pi^{3/2}}=1.162964\dots
\label{aexact}
\eeqa
and
\beqa
C_1&=&1,
\nonumber\\
C_2&=&1,
\nonumber\\
C_3&=&1+\frac{\sqrt{3}}{2\pi}=1.275664\dots,
\nonumber\\
\nonumber\\
C_5&=&1+\frac{5\sqrt{3}\arccos(-1/4)}{2\pi^2}=1.800020\dots,
\label{cexact}
\eeqa
and so
\beqa
B_1&=&1,
\nonumber\\
B_2&=&1-\frac{1}{\pi}=0.681690\dots,
\nonumber\\
B_3&=&1+\frac{2\sqrt{3}-9}{4\pi}=0.559467\dots,
\nonumber\\
B_4&=&1+\frac{\sqrt{3}}{\pi}-\frac{9\arccos(-1/3)^2}{\pi^3}=0.491715\dots,
\label{bexact}
\\
B_5&=&1+\frac{5\sqrt{3}\arccos(-1/4)}{2\pi^2}-\frac{25\arccos(-23/27)^2}{4\pi^3}=0.447534\dots
\nonumber
\eeqa
The above expressions seem to exhaust the list of available closed-form results.
The formulas for $A_2$ and $B_2$ agree with~(\ref{n2sq}) and~(\ref{nvarsq}).

The regime of large dimensions ($D\gg1$) is also worth being investigated.
The leading behavior of the amplitudes $A_D$ and $B_D$ can be derived as follows.
Roughly speaking, the expression~(\ref{phid}) simplifies~to
\beq
\prob{X<x}\sim\exp\left(-D\,\e^{-x^2/2}\right).
\eeq
This estimate expresses that the variable
\beq
\xi=D\,\e^{-X^2/2}
\label{xidef}
\eeq
becomes exponentially distributed,
i.e., that its probability density converges to~$\e^{-\xi}$,
whenever $X$ and $D$ are both large with $\xi$ being finite, i.e.,
\beq
X\approx\sqrt{2\ln D}-\frac{\ln\xi}{\sqrt{2\ln D}}.
\eeq
The random variable $(-\ln\xi)$ entering this expression is a normalized Gumbel variable.
We have in particular $\mean{-\ln\xi}=\gamma$ and $\cum{(\ln\xi)^2}=\pi^2/6$.
We thus obtain the leading-order estimates
\beq
A_D\approx\sqrt{2\ln D},\qquad
B_D\approx\frac{\pi^2}{12\,\ln D}.
\label{abd}
\eeq
Both the exact results~(\ref{aexact}),~(\ref{bexact})
and the estimates~(\ref{abd}) corroborate the observa\-tions made on figure~\ref{fplot}.
The mean value $A_D$ of $X$ slowly increases with $D$, whereas its variance $B_D$ shrinks.
Both effects occur concomitantly at a logarithmically slow pace.

The formulas~(\ref{abd}) are the leading terms
of systematic asymptotic expansions in inverse powers of~$\ln D$
(see~\ref{app} for a self-contained presentation,
as well as~\cite{EVS1}, and~\cite{EVS2,EVS3} for related matters).
The expansions~(\ref{acfull}) yield
\beqa
A_D&=&\sqrt{2\ln D}
\biggl(1+\frac{\nu}{2\ln D}-\frac{6(\nu^2+2\nu+2)+\pi^2}{48(\ln D)^2}
\nonumber\\
&+&\frac{6\nu^3+24\nu^2+3(16+\pi^2)\nu+42+4\pi^2+12\zeta(3)}{96(\ln D)^3}+\cdots\biggr),
\nonumber\\
B_D&=&\frac{\pi^2}{12\ln D}-\frac{\pi^2(\nu+1)+6\zeta(3)}{12(\ln D)^2}+\cdots,
\label{abdsys}
\eeqa
with
\beq
\nu=\gamma-\frac{1}{2}\ln(4\pi\ln D),
\label{nudef}
\eeq
where $\gamma$ is again Euler's constant and $\zeta$ is Riemann's zeta function.

\begin{figure}[!ht]
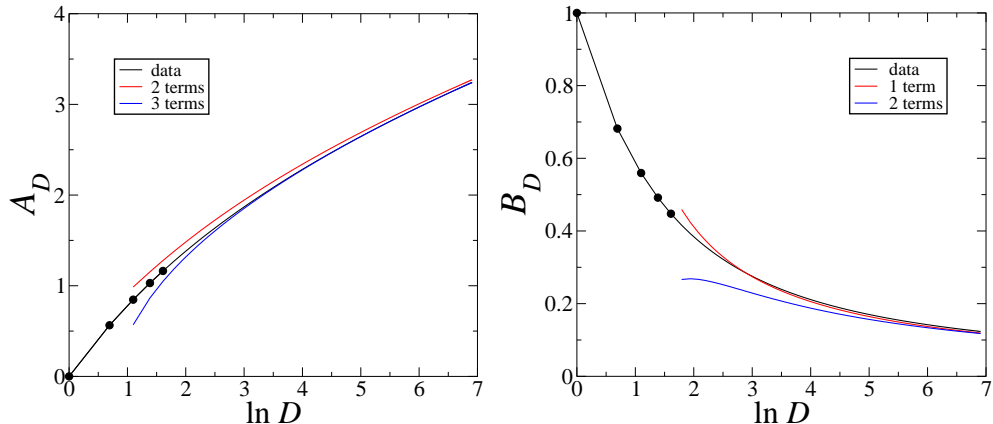

\begin{center}
\includegraphics[angle=0,width=.4825\linewidth]{aplot.eps}
\hskip 3pt
\includegraphics[angle=0,width=.5\linewidth]{bplot.eps}
\caption{\small
The amplitudes $A_D$ (left) and $B_D$ (right) against $\ln D$.
Symbols: exactly known results~(\ref{aexact}),~(\ref{bexact}).
Black curves: numerical results up to $D=1000$.
Red and blue curves: various truncations of the asymptotic expansions~(\ref{abdsys}) (see legends).}
\label{abplot}
\end{center}
\end{figure}

Figure~\ref{abplot} shows the amplitudes $A_D$ and $B_D$ against $\ln D$.
Symbols show the exactly known results~(\ref{aexact}),~(\ref{bexact}).
Black curves show the outcomes of a numerical evaluation
of the integrals entering~(\ref{abcmoms}), up to $D=1000$.
Red and blue curves show various truncations of the asymptotic expansions~(\ref{abdsys}) (see legends).
The amplitude $A_D$ is approached smoothly by the successive approximations,
whereas the situation is more intricate in the case of $B_D$:
data cross the leading term of the asymptotic expansion for $D=19$,
while the second term of that expansion changes sign for $D=3116$, beyond the plotted range.

\section{Summary}
\label{disc}

In this work we have introduced and studied multidimensional record patterns.
These patterns are random sets of lattice points defined by means of a backward recursive construction
generalizing a not-so-well-known but simple and efficient construction
of records in sequences of independent random variables.
The multidimensional construction is elegant in several regards,
as it is parameter-free and respects all symmetries of the underlying lattice.
It was inspired by recent studies
on fragmentation processes of rectangles drawn on the square lattice,
where a rectangle may break either vertically or horizontally
or simultaneously in both directions~\cite{spain,stick}.
The patterns thus generated owe their richness
to the fact that the multidimensional recursive construction is not based on a total order,
except in the one-dimensional situation, where usual records are recovered.

Many exact results have been obtained
on the universal combinatorics of multi\-dimensional record patterns
on finite samples drawn on hypercubic lattices in any dimension $D\ge2$,
generalizing thus standard results of the theory of records.
One remarkable property of the model is that the number $N$ of records
on a sample of size $L_1\times\dots\times L_D$
is exactly distributed as the smallest of the numbers $N_i$ of usual records
in independent signals whose lengths $L_i$ are the sides of the sample.
The most detailed analysis concerns the two-dimensional situation.
We have also investigated the statistics of the landing position $K$,
which is nothing but the length of the stick adjacent to the origin
in the symmetric fragmentation process.
The distribution of $K$ derived here has many common features with
the length distribution of generic sticks in the various fragmentation models
investigated in~\cite{stick},
including multifractal spectra with very similar analytical forms.

In the higher-dimensional situation,
the main focus has been on
asymptotic expressions for quantities of interest on large hypercubic samples,
including the full distribution and the moments of the number of records.
The latter distribution is related to that of the largest of $D$ standard Gaussian variables.
Its mean and variance read
$\mean{N}\approx\ln L-A_D\sqrt{\ln L}$
and $\cum{N^2}\approx B_D\ln L$ (see~(\ref{asymoms})).
The exact values of the amplitudes $A_D$ and $B_D$ have been obtained
up to $D=5$ (see~(\ref{aexact}),~(\ref{bexact})),
whereas asymptotic large-$D$ expansions have also been derived (see~(\ref{abdsys})).

\ack
It is a pleasure to thank Eli Ben-Naim and Sanjay Ramassamy for fruitful discussions,
and Alexander Selby for instructive correspondence.

\appendix

\section{Asymptotic expansions at large dimensions}
\label{app}

The goal of this appendix is to show that
the amplitudes $A_D$, $B_D$ and $C_D$ introduced in~(\ref{abcmoms})
admit systematic asymptotic expansions in inverse powers of $\ln D$.

The leading-order scaling analysis done in section~\ref{hdcubic}
shows that the probability density $f_D(X)$
of the largest of $D$ standard Gaussian variables
gives the highest weight to large positive values of $X$,
such that the combination $\xi$ defined in~(\ref{xidef}) is finite.

This observation can be turned to a systematic analysis as follows.
Setting
\beq
\eps(X)=1-\Phi(X)=\frac{1}{2}\erfc\frac{X}{\sqrt{2}},
\label{eps1}
\eeq
with $\erfc$ being the complementary error function,
the definitions~(\ref{abcmoms}) translate to
\beqa
A_D=\int_0^1D(1-\eps)^{D-1}X(\eps)\dd\eps,
\nonumber\\
C_D=\int_0^1D(1-\eps)^{D-1}X(\eps)^2\dd\eps,
\label{acred}
\eeqa
where $X(\eps)$ is the inverse function of $\eps(X)$.

The first step consists in deriving the asymptotic expansion
of $\eps(X)$ at large $X$ (see e.g.~\cite{GR,AS}).
Setting
\beq
\eps(X)=f(X)g(X)=\frac{\e^{-X^2/2}}{\sqrt{2\pi}}\,g(X),
\label{eps2}
\eeq
with $f(X)=\Phi'(X)=-\eps'(X)$ (see~(\ref{fphi1})),
the second factor obeys the differential equation
\beq
Xg(X)-g'(X)=1,
\eeq
yielding the asymptotic expansion
\beq
g(X)=\frac{1}{X}\sum_{n\ge0}\frac{(2n)!}{n!}\left(-\frac{1}{2X^2}\right)^n.
\eeq

The second step consists in deriving the asymptotic expansion of $X(\eps)$ at small~$\eps$.
From now on we use the shorthand notation $X$ for $X(\eps)$.
Setting
\beq
\lam=\ln\frac{1}{\eps},\qquad\mu=-\frac{1}{2}\ln(4\pi\lam),
\eeq
Equation~(\ref{eps2}) can be recast as
\beq
\frac{X^2}{2}=\lam+\mu-\frac{1}{2}\ln\frac{X^2}{2\lam}+\ln(Xg(X)).
\eeq
Solving this implicit equation iteratively leads to the expansion
\beqa
\frac{X^2}{2}&=&\lam+\mu-\frac{\mu+1}{2\lam}+\frac{2\mu^2+6\mu+7}{8\lam^2}
\nonumber\\
&-&\frac{8\mu^3+42\mu^2+102\mu+107}{48\lam^3}+\cdots,
\label{xasy}
\eeqa
whose structure appears clearly:
the term of order $n$ in $1/\lam$ is a polynomial of degree~$n$ in $\mu$ with rational coefficients.

The third step consists in inserting the expansion~(\ref{xasy}) into the integrals~(\ref{acred}),
after having consistently simplified them to
\beq
A_D\approx\int_0^\infty D\e^{-D\eps}X\,\dd\eps,\qquad
C_D\approx\int_0^\infty D\e^{-D\eps}X^2\,\dd\eps.
\eeq
The latter estimates hold up to negligible corrections scaling as $1/D$,
i.e., exponentially small in $\lam$.
Taking the product $y=D\eps$ as integration variable,
we have $\lam=\ln D-\ln y$, with $\ln D$ being large and $\ln y$ finite.
Using~(\ref{xasy}) to expand $X$ and $X^2$ in inverse powers of $\ln D$,
at every order the integrals over $y$ boil down to the numerical constants
\beq
c_n=\int_0^\infty\e^{-y}(\ln y)^n\dd y,
\eeq
which can be extracted from the following expression of their generating function
\beq
\sum_{n\ge0}\frac{c_ns^n}{n!}=\int_0^\infty\e^{-y}y^s\dd y=\Gamma(s+1),
\eeq
hence
\beqa
c_1&=&-\gamma,
\nonumber\\
c_2&=&\gamma^2+\frac{\pi^2}{6},
\nonumber\\
c_3&=&-\gamma^3-\frac{\gamma\pi^2}{2}-2\zeta(3),
\eeqa
and so on, with $\gamma$ being Euler's constant and $\zeta$ Riemann's zeta function.

Putting everything together, we obtain the expansions
\beqa
A_D&=&\sqrt{2\ln D}
\biggl(1+\frac{\nu}{2\ln D}-\frac{6(\nu^2+2\nu+2)+\pi^2}{48(\ln D)^2}
\nonumber\\
&+&\frac{6\nu^3+24\nu^2+3(16+\pi^2)\nu+42+4\pi^2+12\zeta(3)}{96(\ln D)^3}+\cdots\biggr),
\nonumber\\
C_D&=&2\ln D
\biggl(1+\frac{\nu}{\ln D}-\frac{\nu+1}{2(\ln D)^2}+\frac{6\nu^2+18\nu+21+\pi^2}{24(\ln D)^3}+\cdots\biggr).
\nonumber\\
\label{acfull}
\eeqa
The numerators of the successive terms are polynomials with increasing degrees
in the doubly logarithmic variable
\beq
\nu=\gamma-\frac{1}{2}\ln(4\pi\ln D).
\eeq

\section*{References}

\bibliography{revised.bib}

\end{document}